\documentclass[aps,floatfix]{revtex4}
\usepackage{graphics}
\usepackage{epsfig}
\usepackage{subfigure}
\usepackage{amsmath,amsfonts,amssymb,graphicx,times}

\begin{document}
\title{Opinion percolation in structured population}

\author{Han-Xin Yang$^{1}$}\email{hxyang01@gmail.com}
\author{Liang Huang$^{2}$}\email{huangl@lzu.edu.cn}

\affiliation{$^{1}$Department of Physics, Fuzhou University, Fuzhou
350108, China \\$^{2}$Institute of Computational Physics and Complex
Systems, Lanzhou University, Lanzhou, Gansu 730000, China}

\begin{abstract}
In a recent work [Shao $et$ $al$ 2009 Phys. Rev. Lett. \textbf{108}
018701], a nonconsensus opinion (NCO) model was proposed, where two
opinions can stably coexist by forming clusters of agents holding
the same opinion. The NCO model on lattices and several complex
networks displays a phase transition behavior, which is
characterized by a large spanning cluster of nodes holding the same
opinion appears when the initial fraction of nodes holding this
opinion is above a certain critical value. In the NCO model, each
agent will convert to its opposite opinion if there are more than
half of agents holding the opposite opinion in its neighborhood. In
this paper, we generalize the NCO model by assuming that each agent
will change its opinion if the fraction of agents holding the
opposite opinion in its neighborhood exceeds a threshold $T$ ($T\geq
0.5$). We call this generalized model as the NCOT model. We apply
the NCOT model on different network structures and study the
formation of opinion clusters. We find that the NCOT model on
lattices displays a continuous phase transition. For random graphs
and scale-free networks, the NCOT model shows a discontinuous phase
transition when the threshold is small and the average degree of the
network is large, while in other cases the NCOT model displays a
continuous phase transition.
\end{abstract}

\maketitle



\section{Introduction} \label{sec:intro}

The dynamics of opinion sharing and competing has become an active
topic of recent research in statistical physics~\cite{RMP}. One of
the most successful methodologies used in opinion dynamics is
agent-based modeling~\cite{RMP}. The idea is to construct the
computational devices (known as agents with some properties) and
then simulate them in parallel to model the real phenomena. In
physics this technique can be traced back to Monte Carlo (MC)
simulations~\cite{Landau1}. Beyond relevance as physics models, the
ferromagnetic Ising model~\cite{ising0,ising1,ising2}, the XY
model~\cite{xy} and the Potts model~\cite{Potts1,Potts2} can be seen
as agent-based models for opinion dynamics. Other versions of
opinion models have also been proposed, such as the Sznajd
model~\cite{Sznajd}, the majority rule
model~\cite{majority1,majority2,majority3}, the voter
model~\cite{vote1,vote2}, and the social impact model~\cite{social}.
Some models display a disorder-order
transition~\cite{order1,order2,order3,order4,order5,order6,order7,order8,order9,order10},
from a regime in which opinions are arbitrarily diverse to one in
which most individuals hold the same opinion. Other models focus the
emergence of a global consensus, in which all agents finally share
the same
opinion~\cite{consensus1,consensus2,consensus3,consensus4,consensus5,consensus6,consensus6.1,consensus7,consensus7.1,consensus8,consensus9}.

It has been known that the formation of opinion clusters plays an
important role in opinion dynamics~\cite{op1,op2,op3,op4}. An
opinion cluster is defined as a connected component (subgraph) fully
occupied by nodes holding the same opinion. Recently, Shao $et$
$al$. proposed a nonconsensus opinion (NCO) model~\cite{NCO} in
which each node adopts the majority opinion in its neighborhood at
each time step. It was found that a large spanning cluster of nodes
holding the same opinion appears when the initial fraction of nodes
holding this opinion exceeds a certain threshold~\cite{NCO,li}.
Motivated by the NCO model, Li $et$ $al$. proposed an inflexible
contrarian opinion (ICO) model in which some agents never change
their original opinion but may influence the opinions of
others~\cite{ICO}. It was found that the threshold above which a
large spanning cluster appears is increased with the fraction of
inflexible contrarians.

In both the NCO and ICO models, an agent will adopt the opinion that
is held by more than half of neighbors. However, in many real-life
situations, a quorum far larger than one half is necessary to pass a
resolution. For example, a referendum to recall the president of the
United States requires the support of two-thirds of the senators.
Based on the above reasons, in this paper we generalize the NCO
model by assuming that an agent will change its opinion when the
fraction of agents holding the opposite opinion in its neighborhood
exceeds a threshold $T \geq 0.5$. We call this generalized model as
the NCOT model. When the threshold $T=0.5$, the NCOT model recovers
to the NCO model. When $T=1$, the NCOT model becomes the standard
percolation without opinion dynamics. Both the NCO and ICO models
focus on the critical value for finite-size networks. By the
standard finite-size scaling approach, we have obtained a critical
point at which the phase transition takes place in the limit of
infinite network size. It is interesting to find that, continuous or
discontinuous phase transitions can arise in the NCOT model,
depending on the value of $T$ and the network structure.

The paper is organized as follows. In Sec.~\ref{sec:methods}, we
introduce the NCOT model. In Sec.~\ref{sec:main results}, we study
the NCOT model on square lattices, random networks and scale-free
networks, respectively. Finally, conclusions and discussions are
presented in Sec.~\ref{sec:discussion}.

\section{A nonconsensus opinion model with the threshold (NCOT)} \label{sec:methods}

In the NCOT model on networks, each node holds one of the binary
opinions denoted by +1 and $-1$. Initially, a fraction $f$ of nodes
with the opinion +1 and $1-f$ with the opinion $-1$ are selected at
random. The neighborhood of node $i$ is composed of node $i$ and its
nearest neighbors.  At each time step, each node will convert to its
opposite opinion, if the fraction of nodes holding the opposite
opinion in its neighborhood exceeds a threshold $T$ ($T\geq0.5$).
The system is considered to reach a stable state if no more changes
occur.

\section{Main results} \label{sec:main results}

We focus on the formation of opinion clusters in the NCOT model. We
denote by $S_{1}$ the size of the largest +1 cluster and $S_{2}$ the
size of the second largest +1 cluster in the steady state. Then we
define $s_{1}= S_{1}/N$ and $s_{2}= S_{2}/N$, where $N$ is the
network size. In the following, we carry out simulations
systematically by employing the NCOT model on square lattices,
random networks and scale-free networks respectively.

\subsection{The NCOT model on square lattices}

In this subsection, we study the NCOT model on an $N=L \times L$
square lattice with periodic boundary conditions. Our extensive
numerical simulations reveal that the phase transition of the NCOT
model on square lattice can be roughly divided into three regimes
(different universality classes of percolation) by tuning the
parameter $T$, namely, $T\in [0.5, 0.6)$, $[0.6, 0.8)$ and $[0.8,
1]$. In the same regime, the results are insensitive to the values
of $T$. Particularly, it is noted that for $0.5 \leq T < 0.6$, the
corresponding phase transition pertains to the class of invasion
percolation with trapping~\cite{NCO}; while for $0.8 \leq T \leq 1$,
it gives rise to a phase transition subjecting to regular site
percolation. The percolation threshold is $f_{c}\simeq0.506$ for
$T\in [0.5, 0.6)$~\cite{NCO} and $f_{c}\simeq0.5927$ for $T\in [0.8,
1]$~\cite{Newman}, respectively. For $0.6 \leq T < 0.8$, the phase
transition has some interesting features and has not been reported
before, which will be a focus for this section. To be specific, in
the following simulation, we choose $T=0.7$.

\begin{figure}
\begin{center}
\includegraphics[width=90mm]{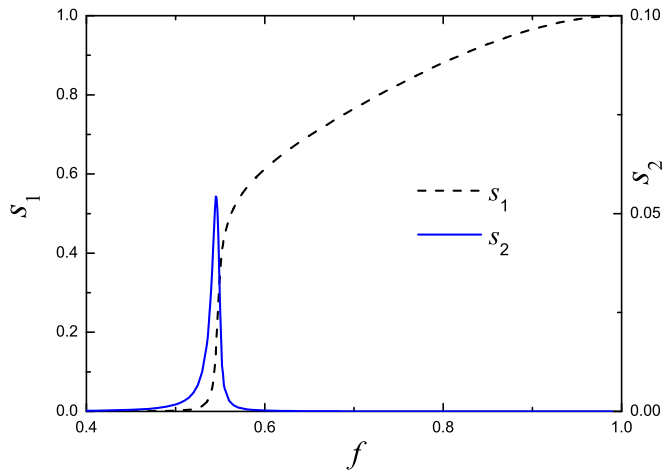}
\caption{The normalized size of the largest cluster $s_{1}$ and the
second largest cluster $s_{2}$ as a function of $f$ on a
$1000\ast1000$ square lattice. $T=0.7$. Each curve is an average of
1000 different realizations.}\label{fig1}
\end{center}
\end{figure}

\begin{figure}
\begin{center}
\includegraphics[width=90mm]{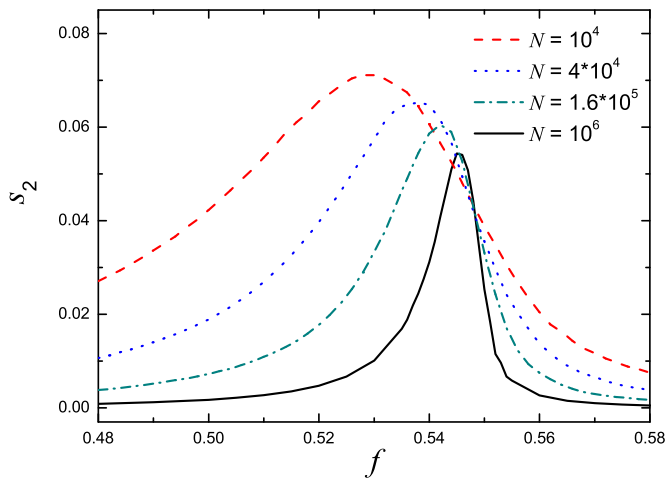}
\caption{The normalized size of the second largest cluster $s_{2}$
as a function of $f$ on a square lattice with different values of
$L$. $T=0.7$. Each curve is an average of 10000, 5000, 3000 and 1000
realizations for $L=100$, 200, 400 and 1000,
respectively.}\label{fig2}
\end{center}
\end{figure}

\begin{figure*}
\begin{center}
\includegraphics[width=150mm]{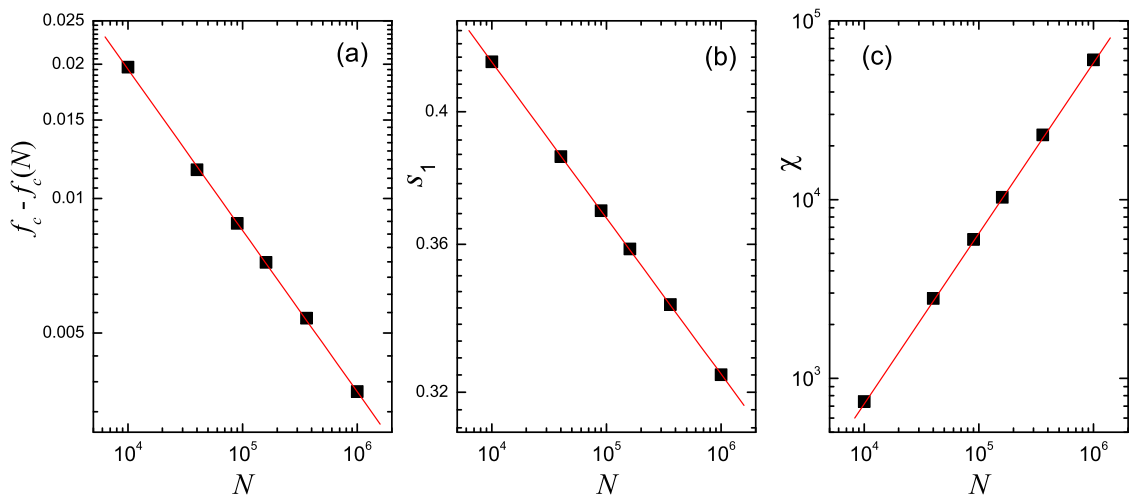}
\caption{(Color online) Log-log plot of (a) $f_{c}-f_{c}(N)$, (b)
the normalized size of the largest cluster $s_{1}$ and (c) the
susceptibility $\chi$, as a function of the system size $N$,
respectively. The percolation threshold $f_{c}\simeq0.5492$ for
$T=0.7$. In (a)-(c), The slopes of fitted lines are -0.36(1),
-0.054(1) and 0.955(8) respectively. Each data point is an average
of 10000, 5000, 4000, 3000, 2000 and 1000 realizations for $L=100$,
200, 300, 400, 600 and 1000, respectively.
 }\label{fig3}
\end{center}
\end{figure*}

Figure~\ref{fig1} shows the normalized size of the largest cluster
$s_{1}$ and the second largest cluster $s_{2}$ as a function of $f$
when $T=0.7$ and $N=10^{6}$. We find that there exists a critical
value $f_{c}(N)$, below which $s_{1}$ approaches 0 and above which
$s_{1}$ continuously increases as $f$ increases. At the critical
value $f_{c}(N)$, $s_{2}$ displays a sharp peak, a characteristic of
a second-order phase transition~\cite{NCO}. Figure~\ref{fig2} shows
the normalized size of the second largest cluster $s_{2}$ as a
function of $f$ for different values of $N$. From Fig.~\ref{fig2},
we observe that the location of $f_{c}(N)$ changes with $N$. The
percolation threshold $f_{c}(N)$ of a system of finite size $N$
obeys the relation~\cite{threshold}
\begin{equation}
f_{c}(N)-f_{c}\sim N^{-1/\nu},
\end{equation}
where $f_{c}$ is the percolation threshold for a system of infinite
size and $\nu$ is the correlation critical exponent. Then a simple
linear fit (based on the maximization of the Pearson's correlation
coefficient) of $f_{c}(N)$ vs. $N^{-1/\nu}$ allows to simultaneously
compute both values of $f_{c}$ and $\nu$~\cite{RRL,PRE}. At the
percolation threshold $f_{c}$, the normalized size of the largest
cluster $s_{1}$ and the susceptibility $\chi=N\sqrt{\langle
s_{1}^{2}\rangle-\langle s_{1} \rangle^{2}}$ vs. the system size $N$
follow a power-law form: $s_{1}\sim N^{-\beta/\nu}$ and $\chi\sim
N^{\gamma/\nu}$.

From simulation results, we obtain $f_{c}\simeq0.5492$ for $T=0.7$.
Figure~\ref{fig3} shows $f_{c}-f_{c}(N)$, $s_{1}$ and $\chi$ as a
function of $N$ respectively. From Figs.~\ref{fig3}(a)-(c), we
estimate the critical exponents $1/\nu=0.36(1)$,
$\beta/\nu=0.054(1)$ and $\gamma/\nu=0.955(8)$ as the best fit of
the data points.

For square lattice, our numerical simulations reveal that the
critical point $f_{c}$ increases as $T$ becomes larger. This is
because for square lattice, the critical point $f_c$ is typically
larger than 0.5. Thus in this region, a larger $T$ hinders the
transition of opinion $-1$ to $+1$ as it requires more $+1$
neighbors, reducing the final fraction of $+1$ nodes in the steady
state, which in turn, will need a larger $f$ for the spanning
cluster of opinion $+1$ to emerge.

\subsection{The NCOT model on random networks}

\begin{figure}
\begin{center}
\includegraphics[width=90mm]{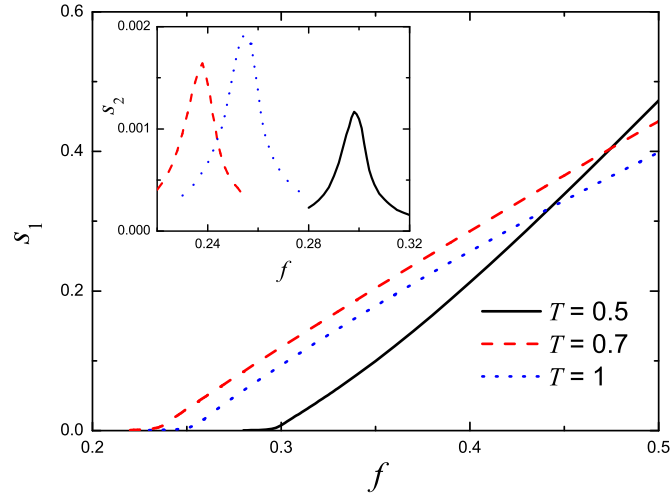}
\caption{The normalized size of the largest cluster $s_{1}$ as a
function of $f$ for different values of $T$. The insets shows the
normalized size of the second largest cluster $s_{2}$ versus $f$ for
different values of $T$. The ER network size $N=10^{6}$ and the
average degree $\langle k \rangle=4$.}\label{fig4}
\end{center}
\end{figure}

\begin{figure}
\begin{center}
\includegraphics[width=80mm]{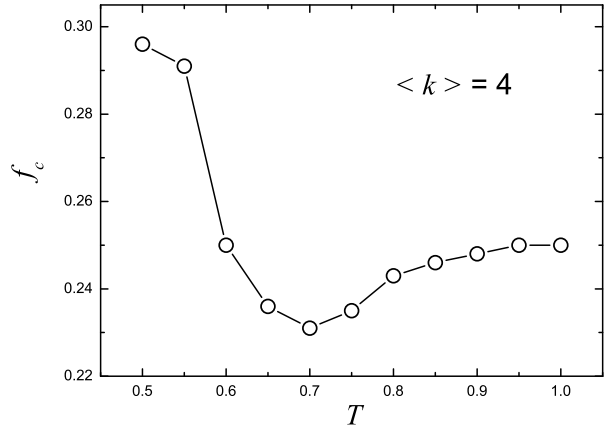}
\caption{The percolation threshold $f_{c}$ as a function of $T$ on
ER networks with the average degree $\langle k \rangle=4$.
}\label{fig5}
\end{center}
\end{figure}

In this subsection, we study the NCOT model on Erd\"{o}s-R\'{e}nyi
(ER) random networks~\cite{er}. ER networks are characterized by a
Poisson degree distribution with $P(k) = e^{-\langle k
\rangle}\langle k \rangle^{k}/k!$, where $k$ is the degree of a node
and $\langle k \rangle$ is the average degree of the network. We
perform simulations with different network sizes $N$. Each data
point presented below is an average over 10000, 8000, 6000, 4000,
3000, 2000 and 1000 different realizations for $N=10^{4}$,
$2\times10^{4}$, $5\times10^{4}$, $10^{5}$, $2\times10^{5}$,
$5\times10^{5}$ and $10^{6}$, respectively.

We have found that, when the average degree is small, the phase
transition is continuous, while when the average degree is
adequately large, depending on $T$, the transition can become
discontinuous.

\begin{figure*}
\begin{center}
\includegraphics[width=150mm]{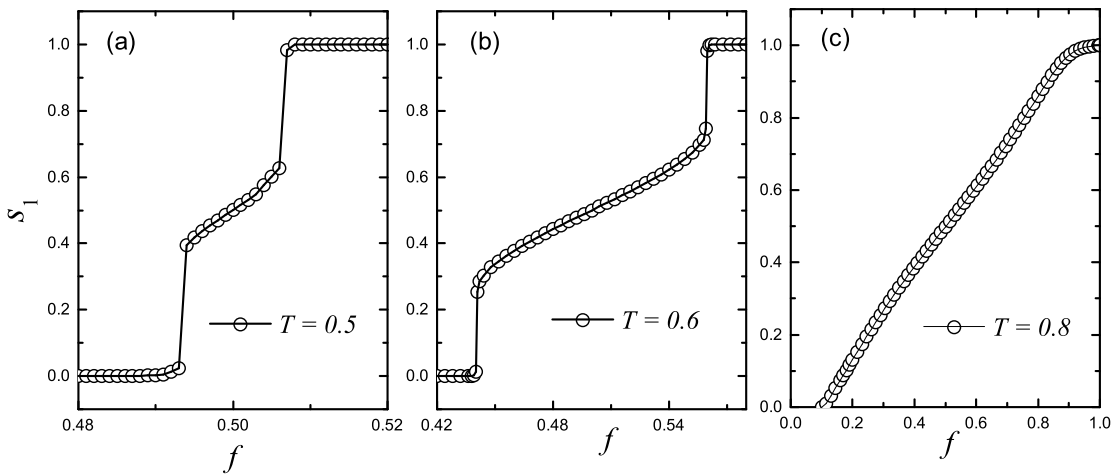}
\caption{The normalized size of the largest cluster $s_{1}$ as a
function of $f$ for (a) $T=0.5$, (b) $T=0.6$ and (c) $T=0.8$. The ER
network size $N=10^{6}$ and the average degree $\langle k
\rangle=10$. }\label{fig6}
\end{center}
\end{figure*}

To be specific, we have found that when the average degree $\langle
k \rangle=4$, the opinion percolation belongs to a continuous phase
transition for the whole range of the parameter $T$. As shown in
Fig.~\ref{fig4}, for different values of $T$ ranging from 0.5 to 1,
the normalized size of the largest cluster $s_{1}$ continuously
increases with $f$ and the normalized size of the second largest
cluster $s_{2}$ peaks at a certain critical value of $f$.
Figure~\ref{fig5} shows the percolation threshold $f_{c}$ as a
function of $T$ when the average degree $\langle k \rangle=4$. The
general trend is that as $T$ increases, the critical threshold fc
decreases. This can be understood that when $T = 0.5$, a node
switches its opinion when half of its neighbors (including itself)
has the opposite opinion, therefore nodes tend to form clusters with
the same opinions. But since $T$ is small, the cluster is not
compact, but rather sparsely connected. In the region where the
initial fraction of nodes with +1 opinion $f < 0.5$, nodes with -1
are majority, the node is more likely to be surrounded by neighbors
with opposite opinions. Thus if $T$ is small, a node with +1 is more
likely to switch to -1, therefore the effective fraction of nodes
with +1 becomes smaller, making it more difficult for a spanning
cluster to emerge. As $T$ increases, it is harder for a node to
switch its opinion (requires more neighbors with opposite opinions),
leading to a larger effective fraction of nodes with +1, and
therefore a higher probability for a spanning cluster to emerge.
Note that the effective fraction of nodes with +1 will still be
smaller than $f$ since nodes with -1 are still majorities therefore
more nodes with +1 will be switched to -1 than the opposite process.

A surprising phenomenon is that there exists an optimal value of $T$
(about 0.7) leading to the minimum of $f_{c}$. This could be a
result of higher order interaction between the network topology and
the opinion dynamics. For example, although a larger $T$ means more
difficult for a node to switch its opinion, but once the condition
is satisfied and this node switches, the cluster grows and it will
be more compact, and it will be more resistive to changes caused by
outside nodes, leading to a nonmonotonic behavior of $f_{c}$ versus
$T$. The nonmonotonic relation between $f_{c}$ and $T$ can also
confirmed in Fig.~\ref{fig4}. From Fig.~\ref{fig4}, we can see that
the critical value of $f$ that corresponds to the peak of $s_{2}$ is
the smallest when $T=0.7$.

When the average degree $\langle k \rangle$ is adequately large
(e.g., $\langle k \rangle=10$), the opinion percolation can display
a continuous or a discontinuous phase transition, depending on the
value of $T$. Figure~\ref{fig6} shows the normalized size of the
largest cluster $s_{1}$ as a function of $f$ for different values of
$T$ when $\langle k \rangle=10$. From Figs.~\ref{fig6}(a) and (b),
we observe that there exist two abrupt transition points when $T$ is
small (e.g., $T=0.5$ and $T=0.6$). At the first abrupt transition
point denoted by $f_{c}$, $s_{1}$ jumps from zero to a finite value.
At the second abrupt transition point denoted by $f_{c}^{\ast}$,
$s_{1}$ jumps from a finite value to one. Between the two abrupt
transition points, $s_{1}$ continuously increases with $f$. However,
for the large value of $T$ (e.g., $T=0.8$), $s_{1}$ approaches zero
continuously as $f$ is decreased from 1 to 0 [see
Fig.~\ref{fig6}(c)]. These numerical results indicate that the
behavior of $s_{1}$ versus $f$ could be a discontinuous phase
transition for the small value of $T$ and a continuous phase
transition for the large value of $T$. From Fig.~\ref{fig6}, it is
seen that the system has a symmetry, i.e., the curves are unchanged
if $f\rightarrow 1-f$ and $s_1 \rightarrow 1-s_1$. This is because,
in this case, the nodes are densely connected, therefore one can
neglect the small clusters for either $+1$ or $-1$ opinions. Denote
$s^*_1$ as the normalized size for the largest cluster for $-1$, we
have $s_1 + s^*_1 \approx 1$. Since the dynamics for the evolution
of $+1$ into $-1$ and vice versa are the same, this imposes a
duality between $+1$ and $-1$ states. Therefore, for a given $f$
after the opinion dynamics reach the steady state the normalized
size of the largest cluster for $+1$ is $s_1$, is actually the same
process for $-1$ opinions with a given initial fraction $1-f$ and a
largest cluster with normalized size $s^*_1(f) \approx 1-s_1(f)$.
But since the dynamics for $+1$ and $-1$ are the same, when the
initial fraction for $+1$ is $1-f$, the largest cluster for $+1$
will also be $s^*_1(f)$, which is approximately $1-s_1(f)$.

\begin{figure}
\begin{center}
\includegraphics[width=120mm]{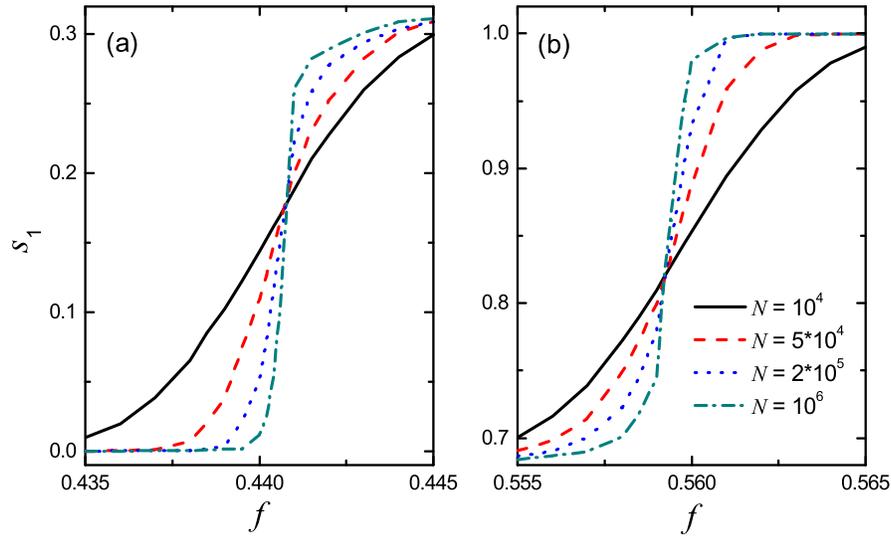}
\caption{The normalized size of the largest cluster $s_{1}$ as a
function of $f$ for different values of the network size $N$. The
average degree of ER networks is $\langle k \rangle=10$ and $T=0.6$.
The initial fraction of nodes with the +1 opinion $f$ is around
$f_{c}$ for (a) and $f$ is around $f_{c}^{\ast}$ for (b).
}\label{fig7}
\end{center}
\end{figure}

\begin{figure}
\begin{center}
\includegraphics[width=85mm]{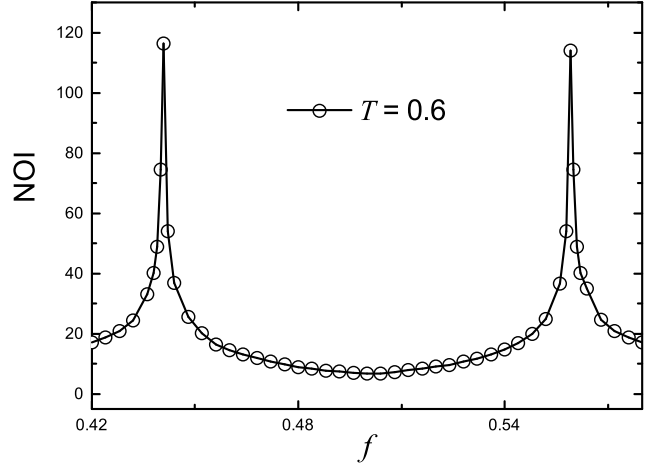}
\caption{ The number of iterations (NOI) as a function of $f$ on ER
networks. The network size $N=10^{6}$, the average degree $\langle k
\rangle=10$ and $T=0.6$. }\label{fig8}
\end{center}
\end{figure}

Figure~\ref{fig7} shows that $s_{1}$ as a function of $f$ for
different values of the network size $N$ when $\langle k \rangle=10$
and $T=0.6$. One can see that all the curves intersect at one point
(the phase transition point). In Fig.~\ref{fig8}, we investigate the
number of iterations (NOI), which is the number of time steps needed
to reach the steady state, as a function of $f$ when $\langle k
\rangle=10$ and $T=0.6$. Note that NOI characterizes the long range
correlation, i.e., if the correlation is local, the system will
quickly settle down to the steady state; while if there exist long
range correlations, it needs more iterations to reach the steady
state since each status change for a node has wider impacts. We can
observe that the NOI exhibits two symmetric peaks. According to
Ref.~\cite{PNAS}, in a discontinuous phase transition, the location
of the peak of the NOI determines the critical threshold of the
transition. In Fig.~\ref{fig8}, the location of the left peak
determines the critical threshold $f_{c}$ below which $s_{1}=0$ and
the right peak determines the critical threshold $f_{c}^{\ast}$
above which $s_{1}=1$. From simulation results shown in Figs. 7 and
8, for $\langle k \rangle=10$ and $T=0.6$, we obtain
$f_{c}\simeq0.4408$ and $f_{c}^{\ast}\simeq0.5592$. It is noted that
$f_{c}+f_{c}^{\ast}=1$, which is consistent with the previous
symmetry analysis.

\begin{figure*}
\begin{center}
\includegraphics[width=150mm]{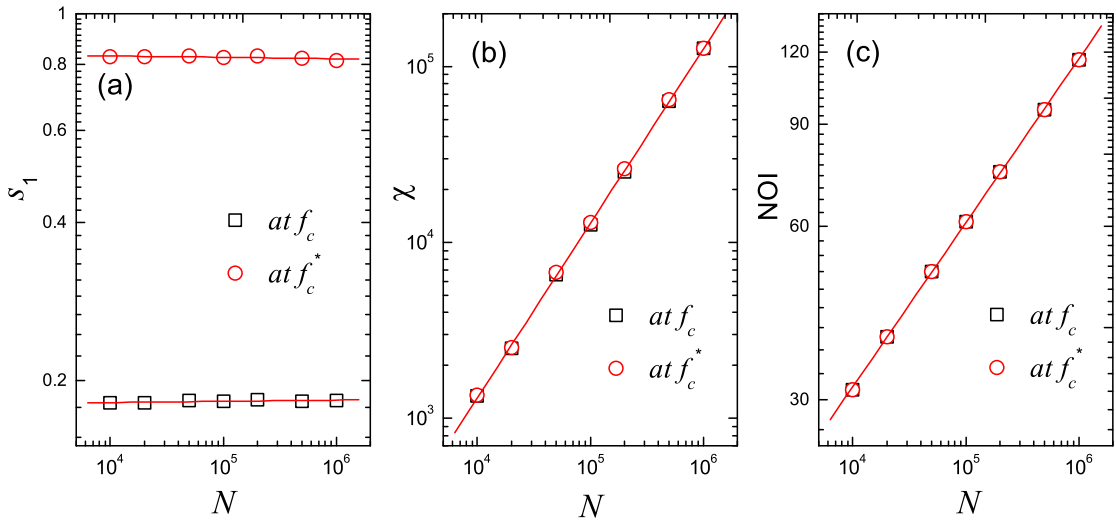}
\caption{Log-log plot of (a) the normalized size of the largest
cluster $s_{1}$, (b) the susceptibility $\chi$ and (c) the number of
iterations (NOI), as a function of the system size $N$,
respectively. In (a)-(c), The slopes of fitted lines are about 0, 1
and 0.28 respectively. The average degree of ER networks $\langle k
\rangle=10$ and $T=0.6$. The phase transition points are
$f_{c}\simeq0.4408$ and $f_{c}^{\ast}\simeq0.5592$. }\label{fig9}
\end{center}
\end{figure*}

To further classify the transition class, we carry out finite size
scaling analysis. Figure~\ref{fig9} shows the normalized size of the
largest cluster $s_{1}$,  the susceptibility $\chi$ and the number
of iterations (NOI), as a function of the system size $N$ at the
discontinuous transition points $f_{c}$ and $f_{c}^{\ast}$,
respectively. Figure~\ref{fig9}(a) shows that $s_{1}$ scales as $
N^{-\beta/\nu}$, with $\beta/\nu\approx0$, indicating a
discontinuous phase transition. Figure~\ref{fig9}(b) illustrates
that $\chi$ scales as $N^{\gamma/\nu}$, with $\gamma/\nu\approx1$.
Figure~\ref{fig9}(c) shows that NOI scales as $N^{\delta}$, with
$\delta\approx0.28$, consistent with the theoretical result $1/4$
\cite{PNAS}.

\begin{figure}
\begin{center}
\includegraphics[width=90mm]{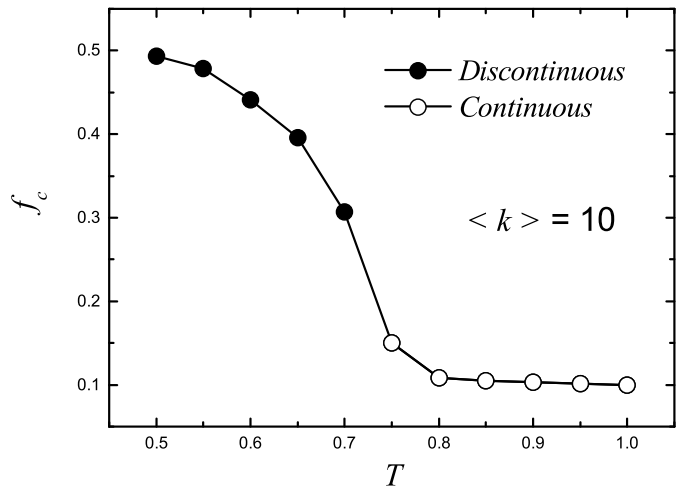}
\caption{The percolation threshold $f_{c}$ as a function of $T$ on
ER networks with the average degree $\langle k \rangle=10$. The
phase transition is discontinuous for small values of $T$ (filled
circles), while it becomes continuous for large values of $T$ (empty
circles).}\label{fig10}
\end{center}
\end{figure}

Figure~\ref{fig10} shows the percolation threshold $f_{c}$ as a
function of $T$ when the average degree $\langle k \rangle=10$. One
can see that $f_{c}$ decreases to 0.1 as $T$ increases. There exists
a certain critical value $T_{c}$ (between 0.7 and 0.75), below which
the phase transition is discontinuous while above which the phase
transition becomes continuous.  The development to the
first-order-like phase transition is originated to the consolidation
of clusters with the same opinions~\cite{zhouhaijun,chenxiaosong}.
Two key factors are needed. The first one is the network topology,
which needs to be dense enough in order to promote clustering
process during the evolution of the opinions~\cite{li}. The second
one is the opinion dynamics, where $T$ should be small to lower the
barrier for a node to switch its opinion to facilitate the formation
of clusters. Therefore, in the region when $\langle k \rangle$ is
large and $T$ is small, one could expect disrupt emerging of
spanning clusters; while in the opposite case when $\langle k
\rangle$ is small and $T$ is large, one would expect continuous
transitions.

\subsection{The NCOT model on scale-free networks}

In this subsection, we study the NCOT model on Barab\'{a}si-Albert
scale-free networks (BA)~\cite{ba}. BA networks are characterized by
a power-law degree distribution with $P(k)\sim k^{-3}$. We perform
simulations with different network sizes $N$. Each data point is an
average over 10000, 6000, 4000, 3000, 2000 and 1000 different
realizations for $N=10^{4}$, $2\times10^{4}$, $5\times10^{4}$,
$10^{5}$, $2\times10^{5}$ and $5\times10^{5}$, respectively.

\begin{figure}
\begin{center}
\includegraphics[width=130mm]{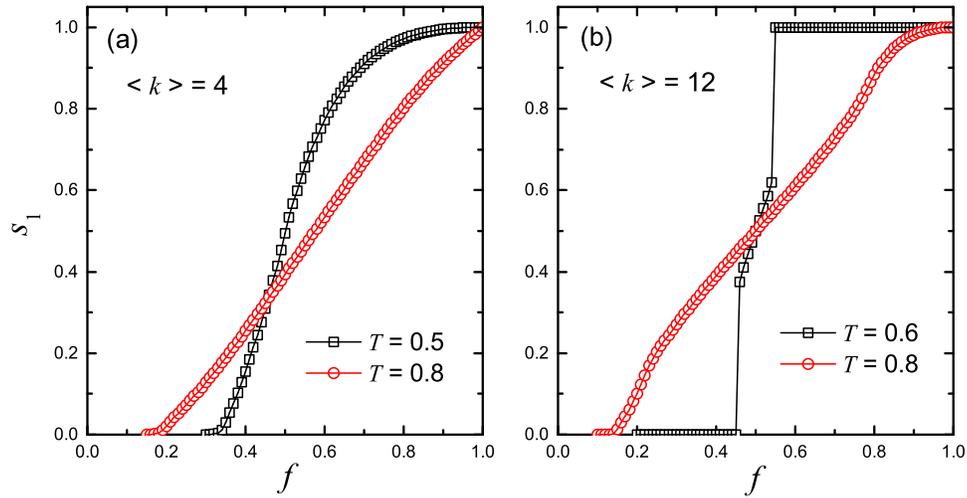}
\caption{The normalized size of the largest cluster $s_{1}$ as a
function of $f$ for different values of $T$. The average degree of
BA networks is (a) $\langle k \rangle=4$ and (b) $\langle k
\rangle=12$, respectively. The network size $N=5\times10^{5}$.
}\label{fig11}
\end{center}
\end{figure}

Figure~\ref{fig11} shows the normalized size of the largest cluster
$s_{1}$ as a function of $f$ for different values of $T$. We can see
that, for $\langle k \rangle=4$ (Fig.~\ref{fig11}(a)), $s_{1}$
approaches zero continuously as $f$ decreases for different values
of $T$, indicating a continuous phase transition. While for $\langle
k \rangle=12$ (Fig.~\ref{fig11}(b)), there exists two abrupt
transition points when $T$ is small (e.g., $T=0.6$), and the
transition becomes continuous for a larger $T$ ($T=0.8$). At the
first abrupt transition point denoted by $f_{c}$, $s_{1}$ jumps from
zero to a finite value. At the second abrupt transition point
denoted by $f_{c}^{\ast}$, $s_{1}$ jumps from a finite value to one.
We have checked that $f_{c}+f_{c}^{\ast}=1$, which is the same as
that in ER networks.

\begin{figure}
\begin{center}
\includegraphics[width=90mm]{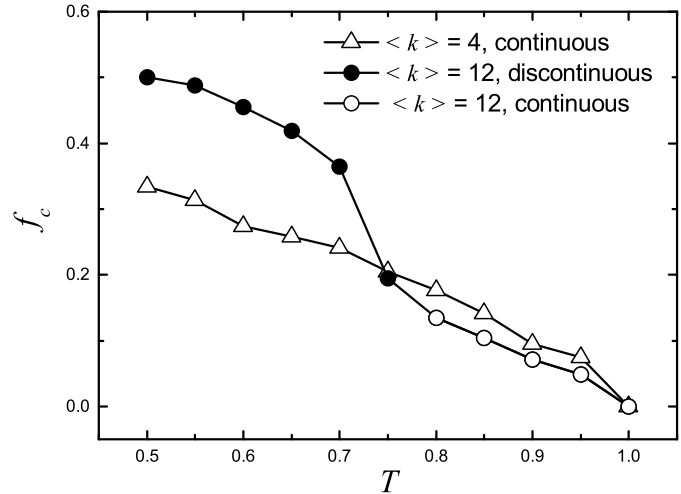}
\caption{The percolation threshold $f_{c}$ as a function of $T$ for
different values of the average degree $\langle k \rangle$ of BA
networks. Filled circles denote that the phase transition is
discontinuous, while empty circles and empty triangles represent
that the transition is continuous.}\label{fig12}
\end{center}
\end{figure}

Figure~\ref{fig12} shows the percolation threshold $f_{c}$ as a
function of $T$ for different values of $\langle k \rangle$. One can
see that $f_{c}$ decreases as the increase of $T$. When the average
degree $\langle k \rangle$ is small (e.g., $\langle k \rangle=4$),
the percolation belongs to a continuous phase transition for all the
values of $T$. When $\langle k \rangle$ is large (e.g., $\langle k
\rangle=12$), the percolation behaves a discontinuous transition for
the small values of $T$ while it displays a continuous phase
transition when $T$ is large.

\section{Conclusions and Discussions} \label{sec:discussion}

In conclusion, we have proposed a generalized nonconsensus opinion
model in which an agent changes its opinion when the fraction of
nodes holding the opposite opinion in its neighborhood exceeds a
threshold $T$ ($T\geq0.5$). We apply the model on various network
structures to study the formation of opinion clusters. It is found
that the behavior of the normalized size of the largest cluster
versus the initial concentration of nodes holding the same opinion
can display a discontinuous or continuous phase transition. For
regular lattices, the phase transition is continuous independent of
$T$. For complex networks such as random networks and scale-free
networks, if the average degree is small, then the phase transition
is continuous, regardless of the value of $T$. For complex networks
with the large average degree, the phase transition is continuous
when $T$ is large but it becomes discontinuous when $T$ is small.
Particulary, there exists two symmetric critical values in the case
of the discontinuous phase transition. The studied opinion
disappears below the first critical value while it takes over the
whole population above the second critical value.

We also study the relationship between $T$ and the percolation
threshold $f_{c}$ above which a large spanning cluster of nodes
holding the same opinion appears. Note that when $T=1$, the phase
transition of our model reverts to the regular site percolation and
the opinion percolation threshold is equal to that of site
percolation. For square lattices, the percolation threshold $f_{c}$
increases as $T$ increases from 0.5 to 1. For Erd\"{o}s-R\'{e}nyi
random networks with the small values of the average degree, there
exists an optimal value of $T$, leading to the minimum $f_{c}$. For
Erd\"{o}s-R\'{e}nyi random networks with the large values of the
average degree or Barab\'{a}si-Albert scale-free networks, $f_{c}$
decreases as the increase of $T$.

\begin{acknowledgments}
We thank Prof. Hai-Jun Zhou and Prof. Wen-Xu Wang for for
constructive comments on the manuscript. This work was supported by
the National Natural Science Foundation of China under grant numbers
61403083, 11422541, 11135001 and 11375074, and the Natural Science
Foundation of Fujian Province of China (Grant No. 2013J05007).
\end{acknowledgments}

\end{document}